\documentclass{article}%
\usepackage{amsmath}
\usepackage{graphicx}
\usepackage{amsfonts}
\usepackage{amssymb}%
\begin{document}

\title{On the spectral shift and the time delay of light in a Rindler accelerated
frame }
\author{J. B. Formiga and C. Romero\\Departamento de F\'{\i}sica, Universidade Federal da Para\'{\i}ba,\\C.Postal 5008, 58051-970 Jo\~{a}o Pessoa, Pb, Brazil\\E-mail: cromero@fisica.ufpb.br }
\maketitle

\begin{abstract}
We discuss two effects predicted by the general theory of relativity in the
context of Rindler accelerated observers: the gravitational spectral shift and
the time delay of light. We show that these effects also appear in a Rindler
frame in the absence of gravitational field, in accordance with the Einstein's
equivalence principle.

\end{abstract}

\section{Introduction}

The concept of Rindler observers is usually taught in special relativity
courses as an example of what is called an uniformly accelerated motion in
Minkowski spacetime \cite{Rindler}. An interesting property exhibited by this
class of accelerated observers, which by definition move rectilinearly with
constant proper acceleration, consists in the fact that if any two of these
observers are connected with a solid rod, then no stresses appear in the rod,
and this is useful to characterize the notion of a "rigid motion" in the
context of special relativity \cite{Rindler}. Apart from these pedagogical
considerations, there has been a great deal of intererest in this kind of
relativistic motion after the theoretical discovery, in the seventies, of the
Unruh effect \cite{Unruh}. As is well known, this is a quantum effect and may
be shortly described as follows. Let us suppose that in a certain inertial
reference frame $S$ some quantum free field $\Phi$, defined in the entire
Minkowski spacetime, is in the vacuum state, i.e. a state in which there is no
particle. Consider now another frame $S^{\prime}$ in which the observers move
with a constant proper acceleration with respect to $S$. What Unruh has showed
is that these accelerated observers will detect not a vacuum state of the
field $\Phi$, but instead a thermal distribution of particles depending on the
proper acceleration of the observers \cite{Unruh}. This means that the concept
of vacuum is relative and depends on the observer.

In spite of its quantum nature, it has been claimed recently that the Unruh
effect has, in fact, a classical (non-quantum) origin due to different
spacetime splittings adopted by the inertial and the accelerated observers
\cite{Pauri,Lachese}. In the present paper we are interested in two
(non-quantum) effects due to accelerated motion that, in principle, might be
measured by Rindler observers: the spectral-shift and the time delay of light.
As we shall see, these two effects are also present in a Rindler frame in the
absence of gravitation and, according to the principle of equivalence, they
might well be interpreted by the accelerated observers as produced by real
gravitational fields \cite{Norton}. Thus the theoretical prediction of the
existence of the two effects constitutes a simple example of Einstein%
\'{}%
s principle of equivalence.

This article is organized as follows. In Sec. I we give a brief review of the
kinematics of Rindler observers. Sections II and III are devoted to the
discussion of the spectral-shift and light time delay as seen by these
accelerated observers. Our conclusions are summarized in Sec. IV.

\section{Rindler accelerated observers}

Let us consider the motion of an uniformly accelerated particle in Minkowski
spacetime $M^{4}$ as viewed from an inertial reference frame $S$. The
worldline of the particle is a timelike curve $\mathcal{C}$ which may be
described by parametric equations $x^{\alpha}=x^{\alpha}(s)$, where $s$ is the
arc length parameter of the curve. (Throughout this work we shall employ the
usual notation $x^{\alpha}=(x^{0}=ct,x^{1}=x,x^{2}=y,x^{3}=z)$, with
$\alpha=0,1,2,3$, and consider Minkowski metric in the form $\eta_{\alpha
\beta}=diag(+---)$). An uniformly accelerated particle is defined, in special
relativity, as the one whose proper acceleration (that is, the acceleration
with respect to its instantaneous rest frame $S^{\prime}$) is constant
\cite{Rindler}. The 4-velocity $u^{\alpha}$ and 4-acceleration $a^{\alpha}$ of
the particle with respect to $S$ are definded as $u^{\alpha}=\frac{dx^{\alpha
}}{ds}$ and $a^{\alpha}=\frac{du^{\alpha}}{ds}$. Because we have chosen to
parametrize $\mathcal{C}$ \ with $s$, the following conditions must hold%
\begin{equation}
u^{\alpha}u_{\alpha}=\eta_{\alpha\beta}u^{\alpha}u^{\beta}=1,
\label{parameter}%
\end{equation}%
\begin{equation}
\text{ }a^{\alpha}u_{\alpha}=\eta_{\alpha\beta}a^{\alpha}u^{\beta}=0.
\label{parameter1}%
\end{equation}
If $a$ denotes the magnitude of the three-dimensional acceleration vector of
the particle with respect to $S^{\prime},$ then we have \cite{footnote1}%
\begin{equation}
a^{\alpha}a_{\alpha}=\eta_{\alpha\beta}a^{\alpha}a^{\beta}=-\frac{a^{2}}%
{c^{4}} \label{acceleration}%
\end{equation}
For simplicity from henceforth we shall restrict ourselves to the
one-dimensional motion in the \ (1+1)-dimensional Minkowski spacetime $M^{2}%
$.The conclusions, however, which will be drawn later\ from this
simplification will naturally hold for the (3+1)-dimensional spacetime
$M^{4}.$ In the (1+1)-dimensional Minkowski spacetime we have $x^{\alpha
}=(x^{0}=ct,x^{1}=x)$ and\ the equations (\ref{parameter}), (\ref{parameter1})
and (\ref{acceleration}) read%
\[
\left(  u^{0}\right)  ^{2}-\left(  u^{1}\right)  ^{2}=1
\]%
\[
a^{0}u^{0}-a^{1}u^{1}=0
\]%
\[
\left(  a^{0}\right)  ^{2}-\left(  a^{1}\right)  ^{2}=-\frac{a^{2}}{c^{4}}%
\]
Since the Eq. (\ref{acceleration}) is the same in all frames of reference
(recall that $a^{\alpha}a_{\alpha}$ is a Lorentz invariant quantity) the
equations above may be used to find the motion of the accelerated particle
with respect to the inertial frame $S$.\ With no loss of generality, let us
consider the motion taking place in the positive direction of the $x$-axis.
Then, from the equations (\ref{parameter}), (\ref{parameter1}) and
(\ref{acceleration})\ we obtain the following system of ordinary differential
equations%
\begin{equation}
\frac{du^{0}}{ds}=\frac{a}{c^{2}}u^{1} \label{eq1}%
\end{equation}%
\begin{equation}
\frac{du^{1}}{ds}=\frac{a}{c^{2}}u^{0} \label{eq2}%
\end{equation}
We can choose the following initial conditions for the motion: $u^{0}(s=0)=1$
and $u^{1}(s=0)=0$, which implies that with respect to $S$, the particle is at
rest $s=0$ \cite{propertime}. Thus we have the solution%
\begin{align}
u^{0}(s)  &  =\frac{dx^{0}}{ds}=\cosh\left(  \frac{as}{c^{2}}\right)
\label{2a}\\
u^{1}(s)  &  =\frac{dx^{1}}{ds}=\sinh\left(  \frac{as}{c^{2}}\right) \nonumber
\end{align}
Further integration\ of the above equations leads to
\begin{align}
x^{0}(s)  &  =\frac{c^{2}}{a}\sinh\left(  \frac{as}{c^{2}}\right)
+b^{0}\label{eq3}\\
x^{1}(s)  &  =\frac{c^{2}}{a}\cosh\left(  \frac{as}{c^{2}}\right)
+b^{1}\nonumber
\end{align}
where $b^{0}$ and $b^{1}$ are integration constants. For convenience we choose
$b^{0}=0$ and $b^{1}=\frac{c^{2}}{a}$ as a second set of initial conditions
(which means that when the accelerated particle is at \ $x=\frac{c^{2}}{a}$
its clock reads $\tau=t=0$, that is, both clocks the particle's and reference
frame $S$\ 's read the same time at $\tau=0$).\ Thus, the equations
(\ref{eq3})\ become%
\begin{equation}
x^{0}(s)=\frac{c^{2}}{a}\sinh\left(  \frac{a\tau}{c}\right)  \label{hyp1}%
\end{equation}%
\begin{equation}
x^{1}(s)=\frac{c^{2}}{a}\cosh\left(  \frac{a\tau}{c}\right)  \label{hyp2}%
\end{equation}
where we are now using the proper time $\tau$ as a parameter\ instead of
$s$.\cite{propertime} Let us briefly comment on the type of motion described
by Eqs. (\ref{hyp1}) and (\ref{hyp2}) as seen by the observers of the inertial
reference frame $S$. In a spacetime diagram the worldline described by these
equations corresponds to the hyperbola (see Fig.1)
\begin{equation}
c^{2}t^{2}-x^{2}=\frac{c^{4}}{a^{2}} \label{hyperbola}%
\end{equation}
(This is why uniformly accelerated motion in Minkowski spacetime\ is
customarily referred to as hyperbolic.\cite{Formiga}).%
\begin{figure}
[ptb]
\begin{center}
\includegraphics[
natheight=715.187500pt,
natwidth=1012.562500pt,
height=234pt,
width=330.5pt
]%
{J1S4LH00.bmp}%
\caption{Hyperbolic motion in Minkowski spacetime representing the worldline
of a particle that experiences a fixed acceleration with respect to its
instantaneous rest frame.\ }%
\end{center}
\end{figure}

From \ the spacetime diagram represented in Fig.1 we see that when
$t\rightarrow\infty$ the worldline of the particle approaches the lightcone of
the event at the origin O\ of the coordinate system of $S$. A simple
calculation shows that the velocity of the particle as seen from $S$
approaches $c$ when $t\rightarrow\infty$. Looking at the whole trajectory we
see that the particle comes in decelerating from infinity to the origin O and
then accelerates back towards $x=+\infty$. A further look at the diagram
pictured in Fig.1 leads us to conclude that the right edges of the lightcone
of O acts as an event horizon for the accelerated observer in the sense that
only events that occur in the region $\left\vert ct\right\vert <x,$ $(x>0),$
(known as the \textit{Rindler wedge}) are accessible to the observer. At this
point, let us note that it is possible to continuously fill the entire Rindler
wedge with hyperbolic worldlines that\ correspond to accelerated observers
with different values of the proper acceleration $a$ $(0<a\mathbf{<\infty)}$.
(\ Simply consider the equation (\ref{hyperbola}) for various values of the
parameter $a$.) It is usual to refer to this family of observers as
\textit{Rindler observers}. The frame constituted by Rindler observers (let us
call it $S_{R}$) is the simplest example of a non-inertial frame in Minkowski
spacetime, and due to its non-inertiality we expect, according to Einstein's
principle of equivalence, that observers of $S_{R}$ can be led to think that
they are in the presence of a "gravitational field". At this point let us
recall that two effects that are typically associated with the presence of a
real gravitational field are the spectral shift and the time delay of
light.\ It is precisely in these two effects that we are interested in this
paper, since according to Einstein's principle of equivalence they should also
appear to accelerated observers, in particular to Rindler observers. For this
purpose it is convenient to establish a coordinate system in which the
accelerated observers are at rest. In other words, we want to have coordinates
$(\tau,\xi)$ that give the proper time and the proper distance as measured by
the accelerated observers.

A set of coordinates with the desired properties mentioned above is provided
by the so-called Rindler \ coordinates. We now proceed to define these
coordinates. First, let us mention that the particles associated with the
Rindler frame \ may be regarded as constituting a rod "moving rigidly" ( in a
relativistic sense) in the $x$-direction.\cite{Rindler} Now out of the whole
family of Rindler observers let us choose an observer\ $\mathcal{O}_{a}$\ with
a given proper acceleration $a.$ Let us consider this observer as standing at
one of the points of the rigid rod mentioned above. We may assign to the
position of $\mathcal{O}_{a}$ in the rod the coordinate $\xi$, so then in
$S_{R}$\ a spacetime diagram the observer%
\'{}%
s worldline will correspond to the curve $\xi=0.$ Let us consider another
point $\mathcal{P}$ of the rod located at a proper distance $\xi$ with respect
to $S_{R}.$The point $\mathcal{P}$, which is at rest (with respect to $S_{R}%
$), is accelerating together with the observer $\mathcal{O}_{a}$. Thus at an
arbitrary time $\tau$ the proper coordinates of $\mathcal{O}_{a}$ and
$\mathcal{P}$ are, respectively, $(\tau,0)$ and $(\tau,\xi)$. Let $S^{\prime}$
be an inertial comoving frame at time $\tau$ with coordinates $(ct^{\prime
},x^{\prime})$ in which the $x^{\prime}$-axis of $S^{\prime}$ coincides with
$\mathcal{O}_{a}$'s rigid rod and $x^{\prime}=0$ gives the position of
$\mathcal{O}_{a}$. We also take $t^{\prime}=\tau$. In the coordinates of
$S^{\prime}$. We now define in $S^{\prime}$ the "displacement vector"
$l^{\prime\alpha}$ at time $\tau$ between $\mathcal{O}_{a}$ and $\mathcal{P}$
by $l^{\prime^{\alpha}}=$ $(\tau,\xi)-(\tau,0)=(0,\xi)$. In the frame $S$ the
components of the displacement vector may be obtained from the Lorentz
transformation
\[
l^{\alpha}=\Lambda_{.\beta}^{\alpha}l^{\prime\beta}%
\]
where $\Lambda_{.\beta}^{\alpha}$ $=%
\begin{bmatrix}
\gamma & \beta\gamma\\
\beta\gamma & \gamma
\end{bmatrix}
,\gamma=\left(  1-\frac{v^{2}}{c^{2}}\right)  ^{-1/2},\beta=\frac{v}{c},$ and
$v$ is the velocity $\frac{dx}{dt}$\ of $\mathcal{O}_{a}$ with respect to $S$.
\ Since $\gamma=u^{0}$ and $\beta\gamma=u^{1}$, we have from (\ref{2a})
\begin{equation}
l^{0}=x_{\mathcal{P}}^{0}-x_{\mathcal{O}}^{0}=u^{1}\xi=\xi\sinh\left(
\frac{as}{c^{2}}\right)  \label{lo}%
\end{equation}%
\begin{equation}
l^{1}=x_{\mathcal{P}}^{1}-x_{\mathcal{O}}^{1}=u^{1}\xi=\xi\cosh\left(
\frac{as}{c^{2}}\right)  \label{l1}%
\end{equation}
Since the point $\mathcal{P}$ may be any point of the rigid rod we have just
established the desired transformation between the coordinates of $S$ and that
of $S_{R}$, which by virtue of (\ref{hyp1}) and (\ref{hyp2}) yields%
\[
x^{0}=\left(  \frac{c^{2}+\xi a}{a}\right)  \sinh\left(  \frac{a\tau}%
{c}\right)
\]%
\[
x^{1}=\left(  \frac{c^{2}+\xi a}{a}\right)  \cosh\left(  \frac{a\tau}%
{c}\right)
\]

The coordinates $(\tau,\xi)$ are\ defined in the intervals $-\infty
<\tau<\infty$ and $-\frac{c^{2}}{a}<\xi<\infty$ and it is easily seen that the
line $\xi=$\ $-\frac{c^{2}}{a}$\ \ is a\ horizon line for the accelerated
observers (see Fig.1).

A simple calculation shows that the metric of Minkowski spacetime in the
proper coordinates $(\xi,\tau)$ is given by
\begin{equation}
ds^{2}=c^{2}dt^{2}-dx^{2}=\left(  c+\frac{\xi a}{c}\right)  ^{2}d\tau^{2}%
-d\xi^{2} \label{Rindler metric}%
\end{equation}
It is important to note that for Rindler observers\ the coordinate $\tau$
plays the role of a time coordinate and that the proper time $\Upsilon
$\ measured by an observer at $\xi=const$ is given by
\begin{equation}
d\Upsilon=(1+\frac{\xi a}{c^{2}})d\tau. \label{propertime}%
\end{equation}
Of course the time coordinate and the proper time are the same for our
observer $\mathcal{O}_{a}$ since by definition $\mathcal{O}_{a}$ has $\xi
=0$.\cite{Dahia}\ \ \ \ \ \ \ \ \ \ \ \ \ \ \ \ \ \ \ \ \ \ \ \ \ \ \ \ \ \ \ \ \ \ \ \ \ \ \ \ \ \ \ \ \ \ \ \ \ \ \ \ \ \ \ \ \ \ \ \ \ \ \ \ \ \ \ \ \ \ \ \ \ \ \ \ \ \ \ \ \ \ \ \ \ \ \ \ \ \ \ \ \ \ \ \ \ \ \ \ \ \ \ \ \ \ \ \ \ \ \ \ \ \ \ \ \ \ \ \ 

\section{Spectral-shift and time delay of light}

We now shall investigate two optical phenomena that in the context of the
general theory of relativity are usually associated with the presence of
gravitational fields. These are the \textit{spectral-shift} and the
\textit{time delay of light}. Being essentially geometrical, these effects
have an interpretation in terms of the bending of the spacetime due to the
presence of matter. Nevertheless, they also may appear in flat Minkowski
spacetime as a purely kinematical effect when we have accelerated observers.
This should not surprise us since from the Einstein's principle of equivalence
gravitational fields can be mimicked by non-inertial reference frames. Our aim
in this section\ is to show how Rindler observers would detect and measure the
spectral and time delay of light. As we shall see, this analysis can be easily
carried out with the help of the Rindler coordinates $(\tau,\xi)$ defined in
Sec. II.\ In the following we shall still be working \ in the (1+1)-Minkowski
since the other two spatial dimensions are not relevant to the discussion.\ \ \ \ 

Let us start with the spectral-shift. Consider the following experiment.
Suppose that a ligth signal is sent by a Rindler observer $\mathcal{A}$\ at a
fixed point $\xi_{\mathcal{A}}$ at time $\tau_{\mathcal{A}}^{(1)}$ (with
respect to the frame $S_{R}$), travels along a null geodesic and is received
by another observer $\mathcal{B}$ a the fixed point $\xi_{\mathcal{B}}$ at
time $\tau_{\mathcal{B}}^{(1)}$.\ Thus the signal passes from the event
\emph{E}$_{1}$ with coordinates $(\tau_{\mathcal{A}}^{(1)},\xi_{\mathcal{A}})$
to the event \emph{R}$_{1}$\emph{ }with coordinates $(\tau_{\mathcal{B}}%
^{(1)},\xi_{\mathcal{B}})$ (see Fig.2).%
\begin{figure}
[ptb]
\begin{center}
\includegraphics[
natheight=715.187500pt,
natwidth=1012.562500pt,
height=234pt,
width=330.5pt
]%
{J1S4LH01.bmp}%
\caption{The worldlines of two Rindler observers who perform an experiment to
detect the spectral shift of light. The light signals travel along null
geodesics from the observer $\mathcal{A}$ to the observer $\mathcal{B}$.}%
\end{center}
\end{figure}

\ Let $\lambda$ be an affine parameter along the null geodesic with
$\lambda=\lambda_{1}$ at the event of the emission and $\lambda=\lambda_{2}$
at the event of reception. Since we have a null geodesic, it follows from
(\ref{Rindler metric}) that%
\begin{equation}
\left(  c+\frac{\xi a}{c}\right)  ^{2}\left(  \frac{d\tau}{d\lambda}\right)
^{2}-\left(  \frac{d\xi}{d\lambda}\right)  ^{2}=0 \label{null}%
\end{equation}
Because the signal travels from $\mathcal{A}$ to $\mathcal{B}$ and $\ $we are
considering $\ \xi_{\mathcal{A}}<$ $\xi_{\mathcal{B}}$ we have
\begin{equation}
\frac{d\tau}{d\lambda}=\left(  c+\frac{\xi a}{c}\right)  ^{-1}\frac{d\xi
}{d\lambda} \label{spectralshift}%
\end{equation}
\ On integrating this equation we obtain
\[
\tau_{\mathcal{B}}^{(1)}-\tau_{\mathcal{A}}^{(1)}=c\int_{\xi_{\mathcal{A}}%
}^{\xi_{\mathcal{B}}}\frac{d\xi}{c^{2}+a\mathbf{\xi}}%
\]
Suppose now that a second light signal is emitted from $\mathcal{A}$ at time
$\tau_{\mathcal{A}}^{(2)}$ (event \emph{E}$_{2}$) and received by
$\mathcal{B}$ at time $\tau_{\mathcal{B}}^{(2)}$ (event \emph{R}$_{2}$). Since
the integral on the right-hand side of (\ref{spectralshift}) depends only on
the fixed positions of the two observers we must have
\[
\tau_{\mathcal{B}}^{(2)}-\tau_{\mathcal{A}}^{(2)}=\tau_{\mathcal{B}}%
^{(1)}-\tau_{\mathcal{A}}^{(1)}%
\]
which then implies
\[
\Delta t_{\mathcal{B}}=\tau_{\mathcal{B}}^{(2)}-\tau_{\mathcal{B}}^{(1)}%
=\tau_{\mathcal{A}}^{(2)}-\tau_{\mathcal{A}}^{(1)}=\Delta t_{\mathcal{A}}%
\]
In other words, the coordinate time differences $\Delta t_{\mathcal{A}}$ and
$\Delta t_{\mathcal{B}}$ between the emission and reception of the light
signal as measured by $\mathcal{A}$ and $\mathcal{B}$\ are the same. However,
as we have mentioned earlier, the proper time measured by these observers
depends on their position. From (\ref{propertime}) the proper time differences
$\Delta\Upsilon_{\mathcal{A}}$ and $\Delta\Upsilon_{\mathcal{B}}$ must be
given by
\[
\Delta\Upsilon_{\mathcal{A}}=(1+\frac{\xi_{\mathcal{A}}a}{c^{2}})\Delta
t_{\mathcal{A}}%
\]%
\[
\Delta\Upsilon_{\mathcal{B}}=(1+\frac{\xi_{\mathcal{B}}a}{c^{2}})\Delta
t_{\mathcal{B}}%
\]
Since $\Delta t_{\mathcal{A}}$ $=$ $\Delta t_{\mathcal{B}}$, we have%
\begin{equation}
\frac{\Delta\Upsilon_{\mathcal{B}}}{\Delta\Upsilon_{\mathcal{A}}}%
=\frac{(1+\frac{\xi_{\mathcal{B}}a}{c^{2}})}{(1+\frac{\xi_{\mathcal{A}}%
a}{c^{2}})} \label{spectral}%
\end{equation}
This is the equation that leads to the spectral-shift formula. Indeed, suppose
our light signal consists of $n$ waves of frequency $\nu_{\mathcal{A}}$, which
are emitted in proper time $\Delta\Upsilon_{\mathcal{A}}$ from the observer
$\mathcal{A}$. Then $n=$\ $\nu_{\mathcal{A}}\Delta\Upsilon_{\mathcal{A}}$.
Observer $\mathcal{B}$ certainly receives $n$ waves, although the frequency
and time duration of the wave train have changed. Clearly we also have
$n=$\ $\nu_{\mathcal{B}}\Delta\Upsilon_{\mathcal{B}}$, which implies
\begin{equation}
\nu_{\mathcal{B}}=\frac{(1+\frac{\xi_{\mathcal{A}}a}{c^{2}})}{(1+\frac
{\xi_{\mathcal{B}}a}{c^{2}})}\nu_{\mathcal{A}} \label{redshift}%
\end{equation}
As we have considered $\xi_{\mathcal{A}}<\xi_{\mathcal{B}}$ Eq.
(\ref{redshift}) tells us that the frequency of light decreases as it leaves
the observer $\mathcal{A}$\ and when it is received at $\mathcal{B}$\ we see a
shift towards the red end of the spectrum. This interesting effect which, in
principle, could be measured by accelerated observers, is the kinematical
analogue of the gravitational red shift, which has been tested by experiment
and been rather well verified.\cite{Pound} \ \ \ \ \ \ \ \ \ \ \ \ \ \ \ \ \ \ \ \ \ \ \ \ \ \ \ \ \ \ \ \ \ \ \ \ \ \ \ \ \ \ \ \ \ \ \ \ \ \ \ \ \ \ \ \ \ \ \ \ \ \ \ \ \ \ \ \ \ \ \ \ \ \ \ \ \ \ \ \ \ \ \ \ \ \ \ \ \ \ \ \ \ \ \ \ \ \ \ \ \ \ \ \ \ \ \ \ \ \ \ \ \ \ \ \ \ \ \ \ \ \ \ \ \ \ \ \ \ \ \ \ \ \ \ \ \ \ \ \ \ \ \ \ \ \ \ \ \ \ \ \ \ \ \ \ \ \ \ \ \ \ \ \ \ \ \ \ \ \ \ \ \ \ \ \ \ \ \ \ \ \ \ \ \ \ \ \ \ \ \ \ \ \ \ \ \ \ \ \ \ \ \ \ \ \ \ \ \ \ \ \ \ \ \ \ \ \ \ \ \ \ \ \ \ \ \ \ \ \ \ \ \ \ \ \ \ \ \ \ \ \ \ \ \ \ \ \ \ \ \ \ \ \ \ \ \ \ \ \ \ \ \ \ \ \ \ \ \ \ \ \ \ \ \ \ \ \ \ \ \ \ \ \ \ \ \ \ \ \ \ \ \ \ \ \ \ \ \ \ \ \ \ \ \ \ \ \ \ \ \ \ \ \ \ \ \ \ \ \ \ \ \ \ \ \ \ \ \ \ \ \ \ \ \ \ \ \ \ \ \ \ \ \ \ \ \ \ \ \ \ \ \ \ \ \ \ \ \ \ \ \ \ \ \ \ \ \ \ \ \ \ \ \ \ \ \ \ \ \ \ \ \ \ \ \ \ \ \ \ \ \ \ \ \ \ \ \ \ \ \ \ \ \ \ \ \ \ \ \ \ \ \ \ \ \ \ \ \ \ \ \ \ \ \ \ \ \ \ \ \ \ \ \ \ \ \ \ \ \ \ \ \ \ \ \ \ \ \ \ \ \ \ \ \ \ \ \ \ \ \ \ \ \ \ \ \ \ \ \ \ \ \ \ \ \ \ \ \ \ \ \ \ \ \ \ \ \ \ \ \ \ \ \ \ \ \ \ \ \ \ \ \ \ \ \ \ \ \ \ \ \ \ \ \ \ \ \ \ \ \ \ \ \ \ \ \ \ \ \ \ \ \ \ \ \ \ \ \ \ \ \ \ \ \ \ \ \ \ \ \ \ \ \ \ \ \ \ \ \ \ \ \ \ \ \ \ \ \ \ \ \ \ \ \ \ \ \ \ \ \ \ \ \ \ \ \ \ \ \ \ \ \ \ \ \ \ \ \ \ \ \ \ \ \ \ \ \ \ \ \ \ \ \ \ \ \ \ \ \ \ \ \ \ \ \ \ \ \ \ \ \ \ \ \ \ \ \ \ \ \ \ \ \ \ \ \ \ \ \ \ \ \ \ \ \ \ \ \ \ \ \ \ \ \ \ \ \ \ \ \ \ \ \ \ \ \ \ \ \ \ \ \ \ \ \ \ \ \ \ \ \ \ \ \ \ \ \ \ \ \ \ \ \ \ \ \ \ \ \ \ \ \ \ \ \ \ \ \ \ \ \ \ \ \ \ \ \ \ \ \ \ \ \ \ \ \ \ \ \ \ \ \ \ \ \ \ \ \ \ \ \ \ \ \ \ \ \ \ \ \ \ \ \ \ \ \ \ \ \ \ \ \ \ \ \ \ \ \ \ \ \ \ \ \ \ \ \ \ \ \ \ \ \ \ \ \ \ \ \ \ \ \ \ \ \ \ \ \ \ \ \ \ \ \ \ \ \ \ \ \ \ \ \ \ \ \ \ \ \ \ \ \ \ \ \ \ \ \ \ \ \ \ \ \ \ \ \ \ \ \ \ \ \ \ \ \ \ \ \ \ \ \ \ \ \ \ \ \ \ \ \ \ \ \ \ \ \ \ \ \ \ \ \ \ \ \ \ \ \ \ \ \ \ \ \ \ \ \ \ \ \ \ \ \ \ \ \ \ \ \ \ \ \ \ \ \ \ \ \ \ \ \ \ \ \ \ \ \ \ \ \ \ \ \ \ \ \ \ \ \ \ \ \ \ \ \ \ \ \ \ \ \ \ \ \ \ \ \ \ \ \ \ \ \ \ \ \ \ \ \ \ \ \ \ \ \ \ \ \ \ \ \ \ \ \ \ \ \ \ \ \ \ \ \ \ \ \ \ \ \ \ \ \ \ \ \ \ \ \ \ \ \ \ \ \ \ \ \ \ \ \ \ \ \ \ \ \ \ \ \ \ \ \ \ \ \ \ \ \ \ \ \ \ \ \ \ 

Let us now turn our attention to the effect of the time delay of light. We
know that in the presence of a gravitational field Einstein's general theory
of relativity has shown that the travel time of light between two given points
in space is greater than it would be in flat spacetime. This difference is
called the light time delay and it has been very well verified
experimentally.\cite{Shapiro} We shall presently show that, as in the case of
the spectral shift, an analogous effect may be measured by accelerated observers.

Consider, as previously, two observers $\mathcal{A}$ and $\mathcal{B}$ at the
fixed points $\xi_{\mathcal{A}}$ and $\xi_{\mathcal{B}}$, respectively.
Suppose that $\mathcal{A}$ sends a light signal towards $\mathcal{B}$ and that
this signal is reflected back from $\mathcal{B}$\ to $\mathcal{A}$ (See
Fig.3).%
\begin{figure}
[ptb]
\begin{center}
\includegraphics[
natheight=715.187500pt,
natwidth=1012.562500pt,
height=232.75pt,
width=328.6875pt
]%
{J1S4LH02.bmp}%
\caption{Light signal sent by the observer $\mathcal{A}$ towards the observer
$\mathcal{B}$ and reflected back from $\mathcal{B}$ to $\mathcal{A}$. The
measure of the time lapse between the events of emission and reception of the
signal reveals the existence of time delay of light.}%
\end{center}
\end{figure}

We can now calculate the time lapse between the events of emission and
reception of the signal. Again, because the signal travels along a null
geodesic we have%
\[
\frac{d\tau}{d\lambda}=\pm\left(  c+\frac{\xi a}{c}\right)  ^{-1}\frac{d\xi
}{d\lambda}%
\]
Assuming now that $\xi_{\mathcal{A}}$ $>$ $\xi_{\mathcal{B}}$\ the coordinate
time $\Delta\tau$ for the whole trip will be given by
\[
\Delta\tau=-\int_{\xi_{\mathcal{A}}}^{\xi_{\mathcal{B}}}\left(  c+\frac{\xi
a}{c}\right)  ^{-1}d\xi+\int_{\xi_{\mathcal{B}}}^{\xi_{\mathcal{A}}}\left(
c+\frac{\xi a}{c}\right)  ^{-1}d\xi=2\int_{\xi_{\mathcal{B}}}^{\xi
_{\mathcal{A}}}\left(  c+\frac{\xi a}{c}\right)  ^{-1}d\xi
\]
The proper time lapse between the emission and reception of the light signal
as recorded by $\ $the clock of the observer $\mathcal{A}$ will be given,
according to Eq. (\ref{propertime}), by
\[
\Delta\Upsilon=2(1+\frac{\xi_{\mathcal{A}}a}{c^{2}})\int_{\xi_{\mathcal{B}}%
}^{\xi_{\mathcal{A}}}\left(  c+\frac{\xi a}{c}\right)  ^{-1}d\xi=2\frac{c}%
{a}(1+\frac{\xi_{\mathcal{A}}a}{c^{2}})\ln\frac{c^{2}+a\xi_{\mathcal{A}}%
}{c^{2}+a\xi_{\mathcal{B}}}%
\]
On the other hand because for the Rindler's observers the distance between the
two observers $\mathcal{A}$ and $\mathcal{B}$ is $\Delta l=$ $\xi
_{\mathcal{A}}-$ $\xi_{B}$ they would expect a round-trip time of
\[
\Delta\overline{\Upsilon}=2\frac{\xi_{\mathcal{A}}-\xi_{\mathcal{B}}}{c}%
\]
It can be shown that $\Delta\overline{\Upsilon}<\Delta\Upsilon$ (see Appendix)
and this constitutes the time delay of light. The difference between
$\Delta\overline{\Upsilon}$ and $\Delta\Upsilon$ is caused exclusively by the
acceleration of the Rindler's observers, that is, it wouldn%
\'{}%
t occur in a inertial reference frame. What we have here \ is a purely
kinematical effect, which in a way reproduces the situation that appears when
a real gravitational field is present.

\section{Final remarks}

We can give an idea of the order of magnitude of the redshift predicted by Eq.
(\ref{redshift}) by considering the observers $\mathcal{A}$ and $\mathcal{B}$
as \ corresponding to the worldlines $\xi_{\mathcal{A}}=0$ and\ \ $\xi
_{\mathcal{B}}=10^{3}m$. For an acceleration $a=10$ $m/s^{2}$ a simple
calculation leads to $\frac{\nu_{B}-\nu_{A}}{\nu_{A}}=\frac{\Delta\nu}{\nu
}\cong-10^{-12}$. It seems that in principle this predicted value is perfectly
measurable with current\ measurement instruments. For comparison note that the
gravitational red shift measured by Pound and Rebka experiment using a
vertical distance of $22.5$ $m$ was $\frac{\Delta\nu}{\nu}=$ $-(5.13\pm
0.51)\times10^{-15}.$ \cite{Pound}.

We can also compute numerically the time delay of light for a configuration of
observers such that $\xi_{\mathcal{B}}=0,$ \ $\xi_{\mathcal{A}}=10^{11}m$,
$a=10$ $m/s^{2}$ . In this case we obtain $\Delta\Upsilon-\Delta
\overline{\Upsilon}\cong3.7\times10^{-4}s$. We should note that the
gravitational time delay, measured by Shapiro, between the transmission of
radar pulses towards (Venus or Mercury) and the detection of the echos is of
the order of $\Delta\Upsilon-\Delta\overline{\Upsilon}\cong10^{-4}%
s$.\cite{Shapiro} Let us remark that the distance involved in the measurement
of the gravitational time delay effect is of the order of $10^{11}m$.

We conclude this section with the following general comment. It is well known
that the equivalence principle, which establishes an equivalence between
gravitational fields and accelerated observers, played a very important role
in Einstein's formulation of the general theory of relativity. Indeed, it was
this powerful physical principle that led him to predict the bending of light
by a gravitatinal field and ultimately to the idea of geometrization of
gravitation.\cite{Einstein} \ By considering a class of accelerated observers,
namely Rindler observers, we have given a simple example of the validity and
power of the principle when applied to two classical effects: the spectral
shift and the time delay of light.

\section{Acknowledgement}

C. Romero would like to thank CNPq-FAPESQ (PRONEX) for financial support.

\section{Appendix}

In this appendix we show that $\Delta\overline{\Upsilon}<\Delta\Upsilon$ for
arbitrary values of the coordinates $\xi_{\mathcal{A}}$ $,\xi_{\mathcal{B}}$ ,
provided that $\xi_{\mathcal{A}}>\xi_{\mathcal{B}}.$ \ With this aim let us
define the function $\mathcal{D(}\xi_{\mathcal{A}},\xi_{\mathcal{B}})$ as
\[
\mathcal{D(}\xi_{\mathcal{A}},\xi_{\mathcal{B}})=2\frac{\xi_{\mathcal{A}}%
-\xi_{\mathcal{B}}}{c}-2\frac{c}{a}(1+\frac{\xi_{\mathcal{A}}a}{c^{2}}%
)\ln\frac{c^{2}+a\xi_{\mathcal{A}}}{c^{2}+a\xi_{\mathcal{B}}}%
\]
For an arbitrary fixed value of $\xi_{\mathcal{B}}$ let us examine the
behaviour of this function when we vary $\xi_{\mathcal{A}}$. For this purpose
take the derivative of \ $\mathcal{D(}\xi_{\mathcal{A}},\xi_{\mathcal{B}})$
with respect to $\xi_{\mathcal{A}}$. We have then
\[
\frac{\partial\mathcal{D(}\xi_{\mathcal{A}},\xi_{\mathcal{B}})}{\partial
\xi_{\mathcal{A}}}=-\frac{2}{c}\ln\frac{c^{2}+a\xi_{\mathcal{A}}}{c^{2}%
+a\xi_{\mathcal{B}}}%
\]
As $\xi_{\mathcal{A}}>\xi_{\mathcal{B}}$ then $\frac{\partial\mathcal{D(}%
\xi_{\mathcal{A}},\xi_{\mathcal{B}})}{\partial\xi_{\mathcal{A}}}$ is negative
for all values of $\xi_{\mathcal{A}}$ $,\xi_{\mathcal{B}}$. Now, for
$\xi_{\mathcal{A}}$ $=\xi_{\mathcal{B}}$ we have $\mathcal{D(}\xi
_{\mathcal{B}},\xi_{\mathcal{B}})=0$. Therefore, we conclude that the function
$\mathcal{D(}\xi_{\mathcal{A}},\xi_{\mathcal{B}})$ is negative for any value
of the coordinates $\xi_{\mathcal{A}}$ $,\xi_{\mathcal{B}}$ $\ $%
($\xi_{\mathcal{A}}>\xi_{\mathcal{B}})$; hence the inequality $\Delta
\overline{\Upsilon}<\Delta\Upsilon$ holds.

\bigskip

\subsection{Caption for figures:}

\textbf{Fig.1} Hyperbolic motion in Minkowski spacetime representing the
worldline of a particle that experiences a fixed acceleration with respect to
its instantaneous rest frame.\ 

\textbf{Fig. 2 }The worldlines of two Rindler observers who perform an
experiment to detect the spectral shift of light. The light signals travel
along null geodesics from the observer $\mathcal{A}$ to the observer
$\mathcal{B}$.

\textbf{Fig. 3 }Light signal sent by the observer $\mathcal{A}$ towards the
observer $\mathcal{B}$ and reflected back from $\mathcal{B}$ to $\mathcal{A}$.
The measure of the time lapse between the events of emission and reception of
the signal reveals the existence of time delay of light.


\begin{thebibliography}{99}
\bibitem {footnote1}Because in rest frame $S^{\prime}$ the velocity of
particle $V^{\prime}=0$ the components of the 4-acceleration in this frame are
$a^{\prime\alpha}=(0,0,a/c^{2},0)$. As the $a^{\prime\alpha}a_{\alpha}%
^{\prime}$ is a relativistic invariant we have $a^{\prime\alpha}a_{\alpha
}^{\prime}=a^{\alpha}a_{\alpha}=-\frac{a^{2}}{c^{4}}$.
\end{thebibliography}
\end{document}